\documentstyle{ioplppt}
\begin{document}

\title{Holonomy and gravitomagnetism}

\author{Roy Maartens\dag\,, Bahram Mashhoon\ddag\, and David R.
Matravers\dag}

\address{~}

\address{\dag Relativity and Cosmology Group, School of Computer
Science and Mathematics, Portsmouth University,
Portsmouth~PO1~2EG, Britain}

\address{~}

\address{\ddag Department of Physics and Astronomy, University of
Missouri-Columbia, Columbia, Missouri~65211, USA}


\begin{abstract}

We analyze parallel transport of a vector field around an
equatorial orbit in Kerr and stationary axisymmetric spacetimes
that are reflection symmetric about their equatorial planes. As in
Schwarzschild spacetime, there is a band structure of holonomy
invariance. The new feature introduced by rotation is a shift in
the timelike component of the vector, which is the holonomic
manifestation of the gravitomagnetic clock effect.

\vskip 10pt

\noindent


\end{abstract}

\section{Introduction}

If a vector is parallel transported around an equatorial circle in
the exterior Schwarzschild spacetime, one may expect that the
vector would be unchanged on return to the starting point, given
the spherical symmetry and staticity of the spacetime. However, it
turns out that in general the vector is shifted after a closed
loop (i.e., there is a deficit angle), as shown by Rothman {\it et
al.}~\cite{rem}. This result reflects the nonlocal nature of
holonomy; parallel transport carries an imprint of the curvature
enclosed by the loop. What is also interesting in the results
of~\cite{rem} is that the shift vanishes for $n$ circuits of a
closed loop at radius $r$ if $n$ and $r$ satisfy an appropriate
condition. In other words, there is a band structure of holonomy
invariance in Schwarzschild spacetime.

In the light of these results, a natural question is how this
holonomy is modified by rotation. Rotation introduces a
gravitomagnetic clock effect~\cite{cm,bgt}, whereby co- and
counter-rotating orbital periods differ, and one expects that the
modified holonomy should reflect the existence of this clock
effect. We consider the holonomy around a closed circle in the
equatorial plane in Kerr and stationary axisymmetric spacetimes
that like Kerr are reflection symmetric about their equatorial
planes. As in the Schwarzschild case, there is a band structure of
holonomy invariance. Unlike the Schwarzschild case, there is a
shift in the timelike component of the transported vector. In
Schwarzschild spacetime, parallel transport around a loop leaves
the timelike component invariant. Rotation of the source and the
associated gravitomagnetism lead to timelike holonomy. For a
four-velocity vector, this holonomy is a Lorentz boost, and the
corresponding local observer would measure a time dilation.

Thus in these stationary spacetimes, holonomy is a geometric
counterpart of the gravitomagnetic clock effect; these are two
related nonlocal signatures of the curvature. (It should be
mentioned that a different treatment of holonomy in Kerr geometry
using Wilson loops is given in~\cite{w}.)

\section{Kerr holonomy}

To study the connection between holonomy and gravitomagnetism
associated with the rotation of the source, we first consider the
exterior Kerr spacetime. In Boyer-Lindquist coordinates, the
metric is~\cite{cm}
\begin{equation}
\!\!\!\!\!\!\!\!\!\!\!\!\!\!\!\!\!\!\!\!\!\!\!\!\!\!\!\!\!\!\!\!\!
(g_{\mu\nu})=\left(
\begin{array}{cccc}
-(1-2Mr/\Sigma) &0&0&-2aMr\sin^2\theta/\Sigma\\ 0& \Sigma/\Delta
&0&0\\ 0&0&\Sigma &0\\ -2aMr\sin^2\theta/\Sigma &0&0&
\sin^2\theta[r^2+a^2+2a^2Mr\sin^2\theta/\Sigma]\\
\end{array}
\right)\;\;,
\end{equation}
where
\begin{equation}
\Sigma=r^2+a^2\cos^2\theta\,,~~ \Delta =r^2-2Mr+a^2\,, \label{del}
\end{equation}
and $x^\mu=(x^0,x^i)=(t, r, \theta, \varphi)$, with $0 <\theta <
\pi$ and $0 \leq \varphi < 2\pi$. The inverse metric in the
equatorial plane, $\theta = \pi/2$, is given by
\begin{equation}
\!\!\!\!\!\!\!\!\!\!\!\!\!\!\!\!\!\!\!\!\!\!\!\!\!\!\!\!\!\!\!\!\!
(g^{\mu\nu}\big|_{\theta=\pi/2})=\left(
\begin{array}{cccc}
-(r^3+a^2r+2a^2M)/(r\Delta) &0&0&-2aM/(r\Delta)\\ 0& \Delta/r^2
&0&0\\ 0&0&1/r^2 &0\\ -2aM/(r\Delta) &0&0& (r-2M)/(r\Delta)\\
\end{array}
\right).
\end{equation}
Units are chosen so that $G = c = 1$; the mass is $M$ and $a =
J/M$ is the specific angular momentum of the Kerr source.

Now consider parallel transport around a circular orbit ($r$ and
$t$ constant) in the equatorial plane to define a vector field
$u^\mu(\varphi)$.  The field satisfies
\begin{equation}
{du^\mu\over d\varphi}+\Gamma^\mu{}_{3\nu}u^\nu=0\,,
\end{equation}
on the orbit.  Since $\Gamma^2{}_{3\mu}=0$ in the equatorial
plane, we have ${du^2/ d\varphi}=0$; moreover, since
$\Gamma^\mu{}_{23}=0$ in the equatorial plane as well, we can
consistently set $u^2=0$ in what follows. For the remaining
components:
\begin{eqnarray}
{du^0\over d\varphi} &=& {aM\over\Delta}\left(3+{a^2\over
r^2}\right)u^1 \,,\label{6} \\ {du^1\over d\varphi} &=&
{\Delta\over r^4}\left[aM\,u^0+(r^3-a^2M)u^3\right] \label{7}\,,
\\ {du^3\over d\varphi} &=& {1\over
r^2\Delta}\left(a^2M+2Mr^2-r^3\right)u^1\,. \label{8}
\end{eqnarray}

The first important conclusion following from these equations is
that in the Schwarzschild spacetime, $du^0/d\varphi=0$. Thus, {\em
there is no holonomy effect on the timelike component in the
absence of rotation.} This is consistent with the absence of a
gravitomagnetic clock effect in the Schwarzschild spacetime.

In the general case with $a \neq 0$, the system in
equations~(\ref{6})--(\ref{8}) gives a third-order equation for
$u^0$:
\begin{equation}
{d^3u^0\over d\varphi^3}+{F(r)\over r^4}\, {du^0\over d\varphi}=0
\,, \label{3de}
\end{equation}
where
\begin{equation}
F(r) = r^{4} - 2Mr^{3} - 2a^{2}Mr - a^{2}M^{2}\,.
\end{equation}
The equation $F(r) = 0$ has only one positive root, $r_*$, by
Descartes' rule of signs. To investigate it we write the equation
in the forms
\begin{equation}
F(r_*) = r_*^{2} \Delta(r_*) - a^{2}(r_* + M)^2 =
0\,,\>\>\>r^3_*(r_*-2M)=a^2M(2r_*+M)\,.
\end{equation}
It follows that $F(r)$ does not vanish on the horizons $r=r_{\pm}$
(with $r_{+} > r_{-}$) at $\Delta = 0$ for $a^2<M^2$, and that
$r_*>2M$ in general. Specifically
\begin{eqnarray}
M^{2} > a^{2}>0~~&&\Rightarrow~~ r_{+} <  2M < r_* <
{\textstyle{5\over2}}M \,,\\ M^{2} =  a^{2}~~&&\Rightarrow~~ r_{+}
< r_* = (1 + \sqrt{2})M\,,\\ M^{2} < a^{2}~~&&\Rightarrow~~  r_* >
2M\,.
\end{eqnarray}
Thus $r_*$ lies beyond the ergosphere, and beyond the horizon in
each case except the last, where there is no horizon. In the limit
of vanishing $a/M$, however, $r_*$ approaches the Schwarzschild
horizon, since, for $a^2/M^2\ll 1$,
\begin{equation}\label{8'}
r_*=2M\left[1+\epsilon-{\textstyle{11\over5}}\epsilon^2
+O(\epsilon^3)\right]\,,~~~~ \epsilon={5\over16}\left( {a\over
M}\right)^2\,.
\end{equation}

The root $r_*$ has a further significance, since it is a critical
value for ``geodesic meeting point" (gmp) observers, for
whom~\cite{bjm1,bjm}
\begin{equation}
\left({d\tau\over dt}\right)_{\rm gmp}^2= {r^2F(r)\over
(r^3-Ma^2)^2}\,,
\end{equation}
where $\tau$ is the proper time for gmp observers. Thus gmp
observers exist only for $r>r_*$~\cite{bjm}.\\

It is useful to divide the solution of Eq.~(\ref{3de}) into three
parts.\\

\noindent{\bf (a)} For $r>r_*$, i.e., $F(r) > 0$, the general
solution is
\begin{eqnarray}
u^0(\varphi)&=& C+ A \, \sin[f(r)\varphi]+ B \,
\cos[f(r)\varphi]\,, \label{sol}\\ u^{1}(\varphi)&=&
\frac{f(r)\Delta (r)}{\Gamma(r)} \left\{
A\,\cos\,[f(r)\varphi]-B\, \sin\,[f(r)\varphi] \right\}
\,,\label{sol'}\\ u^3(\varphi)&=&
\left[\frac{1}{a}-\frac{r+M}{\Gamma(r)}\right]
u^{0}(\varphi)+\frac{CF(r)}{(r^3-a^2M)\Gamma(r)}\,, \label{sol''}
\end{eqnarray}
where $A$, $B$ and $C$ are constants determined by initial
conditions and
\begin{equation}\label{f}
f(r)= \frac{\sqrt{F(r)}}{r^{2}}\,,\>\> \Gamma(r) =
aM\left(3+\frac{a^2}{r^2}\right)\,.
\end{equation}
Starting from $\varphi = 0$ the shift $\delta u^{\mu}$ after $n$
closed loops of parallel transport is
\begin{eqnarray}
\delta u^{0} &=& A \, \sin\left[2 \pi nf(r)\right] - 2 B
\,\sin^{2}\left[\pi nf(r)\right]\,,\label{s1}\\ \delta u^{1}&=&
-\frac{f(r)\Delta(r) }{\Gamma(r)}\left\{A\,{\rm sin^2}\,[\pi
nf(r)] + B\,{\rm sin}\,[2\pi nf(r)]\right\}\,,\label{s2}\\ \delta
u^3 &=& \left[\frac{1}{a}-\frac{r+M}{\Gamma(r)}\right]\delta
u^{0}\,. \label{s3}
\end{eqnarray}

Just as in the Schwarzschild geometry~\cite{rem}, there exists a
band structure of holonomy invariance in the Kerr geometry for
$r>r_*$, i.e., there are radius values $r$ for which $n$ circuits
lead to a net zero shift in the vector. The condition for holonomy
invariance in this case is
\begin{equation}
nf(r) = m~~ \Rightarrow~~ F(r)={m^2\over n^2}\,r^4\,,\label{22}
\end{equation}
where $m$ is a positive integer. For $r>r_*$, Eq.~(\ref{22})
implies that $m^2<n^2$, just as in the Schwarzschild case. It
follows from Descartes' rule of signs that Eq.~(\ref{22}) has only
one positive root $r^*$ for $m^2<n^2$. Furthermore, since $F$ is
positive and monotonically increasing for $r>r_*$, this root is
such that $r^*>r_*$. For fixed $m$, there is a minimum $n$ that
results in holonomy invariance; in particular, no such invariance
exists for finite $r$ if $n=m$. For instance, the holonomy around
a constant-time circle of radius $r^*=3M$ vanishes for $n=9$ and
$m=5$ if the Kerr black hole has $a^2/M^2={2\over7}$. Another
example of holonomy invariance for this circle is provided by
$n=9, m=4$ and $a^2/M^2 = {11\over7}$.

We note that as $a^2/M^2\rightarrow 0$, the band structure of
holonomy invariance reduces to that of Schwarzschild geometry
studied in~[1]; in fact, for $a^2/M^2\ll1$ and fixed integers $n$
and $m$, the solution of Eq.~(\ref{22}) is given by
\begin{equation}
r^*=2M\left[\left(1-\frac{m^2}{n^2}\right)^{-1} +
\epsilon\;\left(1-\frac{m^2}{n^2}\right)
\left(1-\frac{m^2}{5n^2}\right)+O(\epsilon^2)\right]\,,
\end{equation}
where $\epsilon$ is given in Eq.~(\ref{8'}).\\

\noindent{\bf (b)} For $r=r_*$, i.e., $F(r) = 0$, the solution of
Eq.~(\ref{3de}) is simply
\begin{eqnarray}
\label{sol2} u^{0} &=& \tilde C + \tilde{A}\varphi +
\tilde{B}\varphi^{2}\,, \\u^1 & =&
\frac{\Delta(r_*)}{\Gamma(r_*)}\left[\tilde{A}+2\tilde{B}\varphi\right]
\,,\label{sol2'}\\ u^3&=&
\left[\frac{1}{a}-\frac{{r_*}+M}{\Gamma(r_*)}\right]\,u^0
+\frac{2{r_*^4}\tilde B}
{({r_*^3}-a^2M)\Gamma(r_*)}\,,\label{sol2''}
\end{eqnarray}
where $\tilde A, \tilde B$ and $\tilde C$ are constants. The shift
in $u^{\mu}$ after $n$ closed loops is
\begin{eqnarray}
\delta u^0 &=& 2\pi n\tilde A+4\pi^2n^2\tilde B\,,\\ \delta u^1
&=& 4\pi n\frac{\tilde B\Delta(r_*) }{\Gamma(r_*)}\,,\\ \delta u^3
&=& \left[\frac{1}{a} - \frac{{r_*}+M}{\Gamma(r_*)}\right]\delta
u^0\,.
\end{eqnarray}
~\\

\noindent{\bf (c)} For $r<r_*$, i.e., $F(r) < 0$, the solution of
Eq.~(\ref{3de}) is
\begin{equation}
u^0(\varphi)= \hat{C}+ \hat{A} \, \sinh[h(r)\varphi]+ \hat{B} \,
\cosh[h(r)\varphi]\,, \label{sol3}
\end{equation}
where $\hat A$, $\hat B$ and $\hat C$ are constants, and
\begin{equation}
h(r)= \frac{\sqrt{-F(r)}}{r^{2}}\,.
\end{equation}
The other components can be obtained in a similar way from
equations~(\ref{sol})-(\ref{sol''}) by letting $f\rightarrow ih.$
The shift in $u^0$ after $n$ closed loops is
\begin{equation}
\delta u^{0} =  \hat{A}\, \sinh\left[2 \pi nh(r)\right] + 2
\hat{B} \, \sinh^{2}\left[\pi nh(r)\right]\,,
\end{equation}
and the other components of the shift can be determined using
equations~(\ref{s2}) and (\ref{s3}).

\section{Stationary axisymmetric holonomy}

It is possible to extend the main results of our analysis in
Sec.~2 to stationary axisymmetric spacetimes that are symmetric
under a reflection about their equatorial planes. Let us therefore
consider a general stationary axisymmetric spacetime such that in
symmetry-adapted coordinates the metric can be written in the form
\begin{equation}
\left( g_{\mu \nu}\right) = \left(
\begin{array}{cccc}
    g_{tt} & 0 & 0 & g_{t \varphi} \\
    0 & g_{rr} & 0 & 0 \\
    0 & 0 & g_{\theta \theta} & 0 \\
    g_{t \varphi} & 0 & 0 & g_{\varphi \varphi}
    \end{array}
     \right)\,,
\end{equation}
and the inverse metric is given by
\begin{equation}
\left( g^{\mu \nu}\right) = \left(
\begin{array}{cccc}
- \Psi^{-1} g_{\varphi \varphi} & 0 & 0 & \Psi^{-1} g_{t \varphi}
\\ 0 & g_{rr}^{-1} & 0 & 0 \\ 0 & 0 & g_{\theta \theta}^{-1} & 0
\\ \Psi^{-1} g_{t \varphi} & 0 & 0 & - \Psi^{-1} g_{tt}
\end{array}
\right),
\end{equation}
where $\Psi \equiv -(g_{tt}g_{\varphi \varphi} - g_{t
\varphi}^{2})$. Note that $\Psi$ reduces, in the case of the Kerr
metric, to $\Delta\sin^{2} \theta$, where $\Delta$ is defined in
Eq.~(\ref{del}).

The spacetime is stationary, so that the vector normal to the
hypersurfaces of constant time $t$ can be expressed as
$\partial_{t} - G \partial_{\varphi}$, where $G = - g^{t
\varphi}/{g^{tt}} = {g_{t \varphi}}/{g_{\varphi\varphi}}$. In
general $G \neq 0$ and so this vector is different from the
timelike Killing vector $\partial_{t}$.

The requirement that the spacetime be reflection symmetric about
the equatorial plane implies that $g_{\mu \nu}(t, r, \theta,
\varphi) = g_{\mu \nu}(t, r, \pi - \theta,\varphi)$. Therefore
$g_{\mu \nu, \theta}= 0$ for $\theta =\pi/2$. In this case, one
can prove that in general there are circular geodesic orbits in
the equatorial plane. In fact, the radial component of the
geodesic equation for circular orbits with $\theta = \pi/2$
reduces to
\begin{equation}\label{36}
g_{\varphi \varphi, r} \left(\frac{d \varphi}{dt} \right)^{2} + 2
g_{t \varphi, r} \left(\frac{d \varphi}{dt} \right) + g_{tt, r} =
0,
\end{equation}
since $\Gamma^{r}_{\alpha \beta}  = -\frac{1}{2}g^{rr}g_{\alpha
\beta, r}$ for $\alpha,\, \beta = t$  or $\varphi$. Thus there are
in general two angular frequencies in this case corresponding to
co- and counter-rotating orbits. The clock effect arises since in
general $g_{t \varphi, r} \neq 0$. (For further details,
see~\cite{bjm}.)

For the spacetime considered here, equations~(\ref{6})--(\ref{8})
describing the parallel transport of an arbitrary vector field
around a circle of fixed $t$ and $r$ in the equatorial plane take
the form
\begin{eqnarray}
\frac{du^{0}}{d \varphi} & = &\left[ \frac{g^{2}_{\varphi
\varphi}}{2 \Psi}G_{,r} \right]_{\theta=\pi/2} u^{1}\,,\label{37}
\\ \frac{d u^{1}}{d \varphi} & = & \left[ \frac{g_{t \varphi,r}}{2
g_{rr}} \right]_{\theta=\pi/2} u^{0} + \left[ \frac{g_{\varphi
\varphi, r}}{2 g_{rr}} \right]_{\theta=\pi/2} u^{3}\,, \\ \frac{d
u^{3}}{d \varphi} & = & \left[ \frac{g_{tt}g_{\varphi \varphi,r} -
g_{t \varphi} g_{t \varphi , r}}{2 \Psi} \right]_{\theta=\pi/2}
u^{1}\,.
\end{eqnarray}
We have used the fact that, as before, $\Gamma^2{}_{3\mu}=0=
\Gamma^\mu{}_{23}$ in the equatorial plane, and we can
consistently set $u^2=0$. This system leads to
\begin{equation}
\frac{d^{2} {\bf X}}{d \varphi^{2}} + {\cal F}^{2} {\bf X} = 0\,,
\end{equation}
where ${\bf X} = (u^{1}, d u^{0}/d \varphi, d u^{3}/d \varphi)$
and
\begin{equation}
{\cal F}^{2}(r) = -\left\{ \frac{1}{4 g_{rr}
\Psi}\left[g_{tt}(g_{\varphi \varphi,r})^{2} - 2 g_{t \varphi}\,
g_{\varphi \varphi,r}\,g_{t \varphi, r} + g_{\varphi \varphi}(g_{t
\varphi, r})^{2} \right]\right\}_{\theta=\pi/2}.
\end{equation}
In the Kerr case, ${\cal F}$ reduces to $f$, defined in
Eq.~(\ref{f}). For ${\cal F}^{2} > 0$ the band structure of
holonomy invariance exists in general, as in the Kerr case.

\section{Conclusion}

For the Kerr spacetime it follows from the cases (a)--(c)
considered in Sec.~2 that the shift in the timelike component
(i.e., temporal holonomy) is nonzero in general; for instance, if
$u^\mu$ is given initially at $\varphi=0$ by $(0,1,0,0,)$, then
after $n$ loops, ${u^0}$ has grown by virtue of the angular
momentum of the Kerr source. For $r>r_*$,
\begin{equation}
u^0\Big|_{\varphi=0}=0 ~\longrightarrow~ u^0\Big|_{\varphi=2\pi n}
={aM(a^2+3r^2)\over \Delta(r)\sqrt{F(r)}}\,\sin\left[2\pi n
f(r)\right]\,.
\end{equation}
The band structure of holonomy invariance extends from the
Schwarzschild case to the Kerr case, but it does not exist for
$r\leq r_*$ in the Kerr case.

These results refer to the holonomy around a constant-time circle
in the equatorial plane of the Kerr spacetime for an arbitrary
vector field. On the other hand, the corresponding gravitomagnetic
clock effect~\cite{cm,bgt} refers to the motion of clocks on {\it
timelike} circular geodesic orbits. Let $t_{+}(t_-)$ be the period
of a clock on a corotating (counter-rotating) circular equatorial
orbit as measured by asymptotically static inertial observers at
infinity; then,
\begin{equation}
{t_+}-{t_-}=4\pi a\,.
\end{equation}
Moreover, the proper periods $\tau_{\pm}$ accumulated by the
clocks in their complete revolutions around the Kerr source are
such that~\cite{cm,bgt}
\begin{equation}
\!\!\!\!\!\!\!\!\!\!\!\!\!\!\!\!\!\!\!\!\!\!\!\!\!\!\!\!\!
\tau_{+}-\tau_{-}\approx 4\pi
a\left[1+\frac{3}{2}\frac{M}{r}+\frac{27}{8}
\frac{M^2}{r^2}+\left(\frac{135}{16} +
\frac{1}{2}\frac{a^2}{M^2}\right)
\frac{M^3}{r^3}+O\left(\frac{M^4}{r^4}\right)\right]\,.
\end{equation}
Hence $\tau_{+}-\tau_{-}\approx 4\pi a$ for $r\gg 2M$. It follows
from these results that there exists a special temporal structure,
characterized by the specific angular momentum $a$, around a
rotating mass.  This temporal structure is responsible for the
fact that the temporal holonomy involving the {\it timelike}
component of the vector field in Kerr spacetime does not vanish in
general. For a four-velocity vector, the temporal holonomy is a
Lorentz boost, and the associated time dilation signals the
existence of the clock effect. The clock effect and the holonomic
Lorentz boost are purely rotational; both vanish in Schwarzschild
spacetime.

This is the key point of the paper, established first for the Kerr
metric and then extended to the stationary axisymmetric case;
circular equatorial holonomy produces a Lorentz boost, and an
equivalent time dilation, which is a signature of the clock
effect. If $g_{t \varphi,r} \neq 0$ in the equatorial plane, then
in general there is a clock effect by Eq.~(\ref{36}). In addition
$u^{0}(\varphi)$ is not constant in general, since $G_{,r}\neq 0$
in Eq.~(\ref{37}), so that there is timelike holonomy.

We have also shown that the band structure of holonomy invariance
recently demonstrated in Schwarzschild spacetime survives in
appropriately modified form if the source rotates.


\vskip 2cm

\section*{References}

\begin{thebibliography}{99}

\bibitem{rem}
Rothman T, Ellis G F R and Murugan J 2001 {\em Class. Quantum
Grav.} {\bf 18} 1217

\bibitem{cm}
Cohen J M and Mashhoon B 1993 {\it Phys. Lett. A} {\bf 181} 353

\bibitem{bgt}
Mashhoon B, Gronwald F and Theiss D S 1999 {\it Ann. Phys., Lpz.}
{\bf 8} 135

\bibitem{w}
Bollini C G, Giambiagi JJ and Tiomno J 1981 {\em Nuovo Cim. Lett.}
{\bf 31} 13

\bibitem{bjm1}
Bini D, Jantzen R T and Mashhoon B 2001 {\em Class. Quantum Grav.}
{\bf 18} 653

\bibitem{bjm}
Bini D, Jantzen R T and Mashhoon B 2001 preprint (submitted to
{\em Class. Quantum Grav.})

\end{thebibliography}
\end{document}